\documentclass[a4paper,twocolumn,english,english,aps]{revtex4}
\usepackage[T1]{fontenc}
\usepackage[latin1]{inputenc}
\usepackage{graphicx}

\makeatletter

\providecommand{\LyX}{L\kern-.1667em\lower.25em\hbox{Y}\kern-.125emX\@}

\usepackage{babel}
\makeatother
\begin{document}

\title{Skewness as a test of the equivalence principle}

\author{Luca Amendola$^{1}$ \& Claudia Quercellini$^{1,2}$ }

\address{$^{1}$INAF/Osservatorio Astronomico di Roma, \\
 Viale Frascati 33, 00040 Monte Porzio Catone (Roma), Italy }

\address{$^{2}$Università di Roma Torvergata, Via della Ricerca Scientifica
2, 00133, Roma, Italy\\
}

\date{\today {}}

\begin{abstract}
The skewness of the large scale distribution of matter has long been
known to be a probe of gravitational clustering. Here we show that
the skewness is also a probe of violation of the equivalence principle
between dark matter and baryons. The predicted level of violation
can be tested with the forecoming data from the Sloan Digital Sky
Survey.
\end{abstract}
\maketitle
The normalized third order moment of the galaxy distribution, or skewness,
is defined as\begin{equation}
S_{3}=\frac{<\delta (x)^{3}>}{(<\delta (x)^{2}>)^{2}}\, ,\label{eq:}\end{equation}
where $\delta (x)$ is the density contrast at the point $x$. Its
value at large (weakly non-linear) scales can be calculated exactly
assuming that structure forms only via gravitational instability:
in a flat universe dominated by matter with Gaussian initial conditions
the well-known result is \cite{peebles} $S_{3}=34/7$. If the density
contrast is smoothed through a window function of typical size $R$
the skewness becomes \cite{fry,berna}\begin{equation}
\hat{S}_{3}=S_{3}+\frac{d\log \sigma ^{2}(R)}{d\log R}\, ,\label{eq:}\end{equation}
where $\sigma ^{2}$ is the variance of the density field smoothed
through the same window function. In a series of papers it has been
shown that $S_{3}$ remains extremely close to 34/7 in dark energy
models \cite{martel,kam,benabed}, in curved spaces \cite{bouchet,catelan},
in brane-induce gravity \cite{multa} and in Brans-Dicke models \cite{gazta},
with deviations that hardly exceed one per cent in the observationally
acceptable range of cosmological parameters. These results have shown
that $S_{3}$ can be considered one of the best probe of the gravitational
instability picture at large scales \cite{review}. Only scenarios
with radically different features predict values of $S_{3}$ that
deviate sensitively from the standard results: non-Gaussian initial
conditions \cite{gaz-fos}, cosmic strings \cite{ave}, Cardassian
cosmologies \cite{key-5,multa}, and modified gravity models based
on Birkhoff's law \cite{sco}.

Since the skewness is such a good test of gravity, it seems interesting
to ask whether it is also a good test of the universality of gravity,
that is of the equivalence principle. In this paper we focus on the
class of possible violations of the equivalence principle in which
the violating force is mediated by a scalar field. In other words,
we investigate the effects on $S_{3}$ of a scalar field coupled to
matter in a species-dependent way. These models, first studied in
\cite{dam}, have been revived in the context of coupled dark energy
(CDE), in which the same scalar field couples to matter and drives
the accelerated expansion \cite{wet,ame3}. 

Let us recapitulate the calculation of the skewness as detailed in
\cite{fry} and recently reviewed in \cite{review}. We start by writing
down the Newtonian equations for a pressureless fluid with density
$\rho ,$ density contrast $\delta $ and dimensionless peculiar velocity
$v_{i}=v_{pec,i}/\mathcal{H}$ , where $\mathcal{H}=aH$ is the conformal
time Hubble function and $a$ the scale factor. Defining the gravitational
potential $\Phi $ and the auxiliary variable $\Delta =\Phi /(4\pi \rho a^{2})$
the Newtonian equations are \begin{eqnarray}
\delta '+\nabla _{i}(1+\delta )v_{i} & = & 0\, ,\label{eq:cons1}\\
v_{i}'+(1+\frac{\mathcal{H}'}{\mathcal{H}})v_{i}+v_{j}\nabla _{j}v_{i} & = & -\frac{3}{2}\Omega _{m}\nabla _{i}\Delta \, ,\label{eq:eulerdelta}
\end{eqnarray}
and the Poisson equation is $\nabla _{i}\nabla _{i}\Delta =\delta $,
where $\nabla _{i}$ derivates with respect to comoving coordinates
and the prime denotes derivation with respect to $\alpha =\log a$.
These equations have to be complemented by the Friedmann equation
for $\mathcal{H}'$ and by the matter conservation equations. 

We generalize now the equations introducing a scalar coupling to dark
matter. Such a coupling is realized in any theory which admits in
the Lagrangian a Brans-Dicke term of the form $f(\tilde{\phi },\tilde{R})$;
the low-energy limit of superstring theory is the most interesting
example \cite{gasp2001}. Upon a conformal transformation, this theory
can be written as Einsteinian gravity in which matter and scalar field
interact through an exchange term in their conservation equations
(see e.g. \cite{wet,ame1,paper1}):\begin{eqnarray}
T_{(c)\nu ;\mu }^{\mu } & = & -\sqrt{2/3}\kappa ^{2}\beta (\phi )T_{(c)}\phi _{;\nu }\, ,\label{eq:cons}\\
T_{(\phi )\nu ;\mu }^{\mu } & = & \sqrt{2/3}\kappa ^{2}\beta (\phi )T_{(c)}\phi _{;\nu }\, ,\nonumber 
\end{eqnarray}
where $\kappa ^{2}=8\pi G$ and the dimensionless coupling $\beta (\phi )$
depends on the function $f(\tilde{\phi },\tilde{R})$. The coupling
introduces two distinct effects on the Newtonian equations: first,
due to the interaction an additional force appears as a source; second,
the matter energy density is no longer conserved. The first effect
implies that matter feels an extra force due to its interaction with
the scalar field that will add to the right-hand-side in Eq. (\ref{eq:eulerdelta}).
We can write this term in all generality as $-\frac{3}{2}\Omega _{m}\eta (a,r)\nabla _{i}\Delta $
where $\eta (a,r)$ is a function that in general will depend on time
and distance. If the scalar potential is $V(\phi )$ then its effective
mass is $m^{2}=d^{2}V/d\phi ^{2}$ and the interaction scale is $\lambda =1/m$.
Then the new force introduces a Yukawa correction in the gravitational
potential which becomes $\eta (r)/r$, where $\eta (r)=1+(4/3)\beta ^{2}e^{-r/\lambda }$.
Here, however, we will assume for simplicity that $\lambda $ is infinite,
or at least much larger than the observed scales: on one hand, a very
small $\lambda $ would be unobservable, since the interaction would
be effectively damped; on the other, if we interpret the scalar field
as dark energy then its interaction scale $\lambda $ is expected
to be of the order of the Hubble size (see \cite{paper1}). It is
possible however to generalize the calculations to a finite $\lambda $
.

The second effect arises because of the non-conservation of the matter
energy density. Eq. (\ref{eq:cons}) implies in fact in a homogeneous
and isotropic metric an equation of the form\begin{equation}
\rho '+3\rho =-\sqrt{2/3}\kappa ^{2}\beta (\phi )\phi '\rho \, ,\label{eq:}\end{equation}
whose solution $\rho =n_{0}m_{0}a^{-3}e^{-\sqrt{2/3}\kappa ^{2}\int \beta (\phi )d\phi }$
can be interpreted as a varying dark matter mass, $\rho =n_{0}a^{-3}m(\phi )$
with $m(\phi )=m_{0}e^{-\sqrt{2/3}\kappa ^{2}\int \beta (\phi )d\phi }$
and $n_{0}$ is the numerical density of particles at present. This
time-dependent mass introduces an extra friction term in the Euler
equations. Therefore, the scalar-Newtonian Euler equation for a coupled
fluid can be written in the form\begin{equation}
\mathbf{v}_{c}'+\frac{1}{2}F_{c}(\alpha )\mathbf{v}_{c}+(\mathbf{v}_{c}\cdot \overrightarrow{\nabla })\mathbf{v}_{c}=-\frac{3}{2}S_{c}(\alpha )\overrightarrow{\nabla }\Delta _{c}\, ,\label{veleq}\end{equation}
where the two functions, the friction $F_{c}(\alpha )$ and the source
$S_{c}(\alpha )$, are in general time-dependent. In \cite{paper1}
we have shown that the full relativistic perturbation treatment of
CDE reduces to Eqs. (\ref{eq:cons1},\ref{veleq}) in the Newtonian
limit, with \begin{eqnarray*}
F_{c}(\alpha ) & = & 2[1+\frac{\mathcal{H}'}{\mathcal{H}}-2\beta \frac{\phi '}{\sqrt{6}}]\, ,\\
S_{c}(\alpha ) & = & \Omega _{c}\left(1+\frac{4}{3}\beta ^{2}\right)\, ,
\end{eqnarray*}
 where the $\beta $ terms in $F_{c},S_{c}$ quantify the two effects
due to the scalar interaction. Let us stress again that the function
$\beta $ is in general field-dependent.

The upper bounds on a scalar interaction with baryons (subscript $b$)
are very strong, of the order of $\beta _{b}<0.01$ \cite{hagi}:
in the following we assume that the interaction to baryons is effectively
zero but will generalize to $\beta _{b}\not =0$ at the end ( in Ref.
\cite{khoury} it has been proposed a model in which such constraints
can be escaped but only for suitably chosen potentials). The bounds
on a coupling to dark matter (subscript $c$) are however much weaker.
In \cite{gra} astrophysical observations were employed to derive
$\beta _{c}<1.5$ roughly; $N$-body simulations \cite{maccio} have
shown that the dark matter halo profile depends sensitively on $\beta _{c}$
in a class of dark energy models but, due to the controversial status
of the halo profile observations, it is difficult to derive firm upper
limits. Finally, in \cite{amen02} we found that CMB requires $\beta _{c}<0.13$;
however, this result assumes that the coupling remains constant throughout
the universe lifetime and it is actually most sensitive to the value
of $\beta _{c}$ at early times. Moreover, the limits obviously depends
on the assumed priors on the cosmological parameters, especially on
the Hubble constant. We can summarize the observational situation
with respect to $\beta _{c}$ by saying that there are no strong upper
bounds to the \emph{present} value of a scalar field coupling to dark
matter; if $\beta _{c}$ varies with time then even a value of order
unity at present is not definitively excluded. We assume therefore
the coupling to dark matter as a free parameter and drop the subscript
$c$ in $\beta _{c}$. Since the baryons are practically uncoupled,
the scalar interaction violates the equivalence principle. The value
of $\beta $ is therefore also a measure of the equivalence principle
violation.

For an uncoupled and subdominant (i.e. $\Omega _{b}\ll \Omega _{c}$)
component like the baryons Eq. (\ref{veleq}) becomes \begin{equation}
\mathbf{v}_{b}'+\frac{1}{2}F_{b}(\alpha )\mathbf{v}_{b}+(\mathbf{v}_{b}\cdot \overrightarrow{\nabla })\mathbf{v}_{b}=-\frac{3}{2}S_{b}(\alpha )\overrightarrow{\nabla }\Delta _{c}\, ,\label{veleqb}\end{equation}
where $F_{b}=2+2\mathcal{H}'/\mathcal{H},\, \, S_{b}=\Omega _{c}+\Omega _{b}\approx \Omega _{c}$.
In the standard pure matter case $\mathcal{H}'/\mathcal{H}=-1/2$
and $F_{c,b}=S_{c,b}=1$, while in a flat universe with a scalar field
component with equation of state $w_{\phi }=p_{\phi }/\rho _{\phi }$
\begin{equation}
\frac{\mathcal{H}'}{\mathcal{H}}=-\frac{1}{2}\left[1+3w_{\phi }(1-\Omega _{m})\right]\, \label{eq:hph}\end{equation}
(here and in the following $\Omega _{m}=\Omega _{c}+\Omega _{b}$). 

Following the notation of refs. \cite{fry} and \cite{kam} we expand
the scalar-Newtonian equations in a perturbation series, $\delta =\sum _{i}\delta ^{(i)},$
and $\Delta =\sum _{i}\Delta ^{(i)}.$ It is convenient to define
for each component the growth function $D_{1}(\alpha )$, \begin{equation}
\delta ^{(1)}=D_{1}(\alpha )\delta _{0}^{(1)},\label{eq:grow}\end{equation}
where $\delta _{0}^{(1)}$ is the density contrast at the initial
time (assumed Gaussian distributed) and the growth exponent $m(\alpha )=D_{1}\, '/D_{1}.$
At first order we derive therefore the equations\begin{eqnarray}
\delta _{b}^{(1)}\, ''+\frac{F_{b}}{2}\delta _{b}^{(1)}\, '-\frac{3}{2}S_{b}\delta _{c}^{(1)} & = & 0\, ,\label{eq:d1b}\\
\delta _{c}^{(1)}\, ''+\frac{F_{c}}{2}\delta _{c}^{(1)}\, '-\frac{3}{2}S_{c}\delta _{c}^{(1)} & = & 0\, .\label{eq:d1c}
\end{eqnarray}
 Asymptotically, the dark matter drives the evolution of the baryons,
so that the two components grow with the same exponent $m(\alpha )$
but with a biased amplitude, $b=\delta _{b}^{(1)}/\delta _{c}^{(1)}$.
Subtracting the two equations we see that $b=S_{b}/[S_{c}+m(F_{b}-F_{c})]$
\cite{paper1}. In the limit of small $\phi '$, which is a typical
occurrence for a slowly-rolling dark energy field, $F_{c}=F_{b}$
and the (anti-)bias is simply\begin{equation}
b=(1+4\beta ^{2}/3)^{-1}\, ,\label{eq:}\end{equation}
 constant in time and space. In the same limit it appears that both
the background evolution and the perturbation equations depend on
$\beta ^{2}$ so that the sign of $\beta $ is irrelevant. In it remarkable
that in the opposite limit in which the field kinetic energy is much
larger than the potential energy (for instance in the original Brans-Dicke
model in which $V=0$) so that the field does not drive the acceleration,
it turns out that $\phi '\propto \beta $ \cite{amen02} and the product
$\beta \phi '$ in $F_{c}$ is proportional to $\beta ^{2}$. Then
even in this case the sign of $\beta $ does not matter. In the following
we put $\beta >0$.

Here and below we will assume for the numerical integrations an inverse
power law potential $V\sim \phi ^{-n}$. The potential appears only
in the background equations and, indirectly, in the assumption $\lambda \to \infty $.
For this potential the present equation of state is approximated by
$w_{\phi 0}=-2/(n+2)$ \cite{zla} during the tracking regime (which
may or may not extend to the present epoch; in the latter case $w_{\phi 0}\to -1$).
Integrating numerically Eqs. (\ref{eq:d1b}-\ref{eq:d1c}) we find
a fit for $m$ \begin{equation}
m\approx \Omega _{m}^{0.56(1-1.73\beta ^{2})}\, ,\label{eq:m}\end{equation}
almost independent of $n$ (we explored the range $n\in (0,2)$).

We proceed now to second order. We define for each component $b,c$
the second-order Fourier amplitude $\delta _{b,c}^{(2)}(k,\alpha )=D_{2b,c}(\alpha )\delta _{b,c}^{(2)}(k)$
and, following the standard technique of Fourier convolution (see
e.g. \cite{fry,kam,review}) we obtain for $D_{2b,c}$ the equations\begin{eqnarray}
D_{2b}^{''}+\frac{F_{b}}{2}D_{2b}^{'}-\frac{3}{2}S_{b}D_{2c} & = & \left(\frac{3}{2}S_{b}b+\frac{4}{3}m^{2}b^{2}\right)D_{1}^{2},\label{eq:d2b}\\
D_{2c}^{''}+\frac{F_{c}}{2}D_{2c}^{'}-\frac{3}{2}S_{c}D_{2c} & = & \left(\frac{3}{2}S_{c}+\frac{4}{3}m^{2}\right)D_{1}^{2}.\label{eq:d2c}
\end{eqnarray}
 with the initial conditions $D_{2b,c}(\alpha _{in})=D_{2b,c}'(\alpha _{in})=0$.
It is interesting to note that equations similar to (\ref{eq:d2c},\ref{eq:d1c})
have been derived in \cite{multa,sco} for a general single-component
density expansion of Friedmann equations. However, the mass non-conservation
induced by scalar gravity introduces non-standard friction terms $F_{b,c}$
that cannot be accounted for within the class explored in Ref. \cite{multa,sco}.

As it has been shown in \cite{kam,benabed}, the dominant term in
the skewness is\[
S_{3c}=6\frac{D_{2c}}{D_{1}^{2}}\, ,\quad S_{3b}=6\frac{D_{2b}}{b^{2}D_{1}^{2}}\, .\]
To derive an approximate analytical solution we can assume $D_{2b}=b^{(2)}D_{2c}$
with a constant second-order bias $b^{(2)}$since, as in the linear
equations, the baryon evolution is driven by the dark matter one.
The bias $b^{(2)}$ is not to be confused with the bias $b_{2}$ employed
in literature in the Taylor expansion of a non-linear mapping from
the underlying matter density to the galaxy distribution, see e.g.
\cite{review}. It turns out that for small $\beta ^{2}$ the leading
non-trivial term is\begin{equation}
\frac{S_{3b}}{S_{3c}}=\frac{b^{(2)}}{b^{2}}\approx 1+\beta ^{2}\left(\frac{34\Omega _{m}}{28m^{2}+57\Omega _{m}}\right)\, .\label{eq:s3bappr}\end{equation}
 (here we employed the approximation $S_{3c}\approx 34/7,$ see below).
Substituting (\ref{eq:m}) it appears that $S_{3b}$ is almost independent
of the potential slope $n$ and, since $m^{2}\approx \Omega _{m}$,
also almost independent of $\Omega _{m}$.

\begin{figure*}
\begin{center}\includegraphics[  bb=0bp 200bp 462bp 742bp,
  clip,
  scale=0.6]{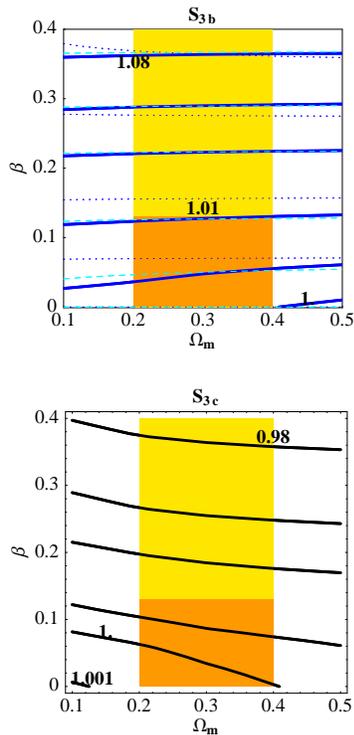}\end{center}

\caption{Contour plot of the observable quantities $S_{3b}\cdot (7/34),S_{3c}\cdot (7/34)$
calculated numerically as function of $\Omega _{m},\beta $. For $S_{3b}$
the lines correspond to the contour values $1.,1.002,1.01,1.03,1.05,1.08$,
while for $S_{3c}$ they are $1,1.001,0.999,0.995,0.99,0.98$, both
from bottom to top. The short-dashed curves in the plot $S_{3b}$
are the fit (\ref{eq:fits3b}); the dotted curves the approximation
(\ref{eq:s3bappr}). The light rectangle marks the astrophysical bounds
on $\Omega _{m}$, the darker one adds the CMB constraint assuming
constant $\beta $ .}
\end{figure*}

The system (\ref{eq:d1b}-\ref{eq:d1c}-\ref{eq:d2b}-\ref{eq:d2c}),
along with the background equations (\ref{eq:cons}) and (\ref{eq:hph}),
constitute a complete set of differential equations for the unknowns
$b,m,S_{3b},S_{3c}$ as function of the cosmological parameters $w_{\phi },\Omega _{m},\beta $.
Each of the functions $b,m,S_{3b},S_{3c}$ depends on $\beta $ and
is therefore in principle a test of the equivalence principle and,
more in general, of coupled dark energy. In the long-term this redundancy
can be exploited to set more stringent limits to the coupling and
to break degeneracies with other cosmological parameters. However,
$b$ and $S_{3c}$ require the detection of the large scale clustering
of the dark matter component, while $m$, the clustering growth rate,
requires accurate observations over an extended range of redshifts
and, consequently, the problematic removal of redshift-dependent selection
effects. Moreover, Eq. (\ref{eq:m}) implies a strong level of degeneracy
between $\beta $ and the parameters that enter $\Omega _{m}(z)$.
On the other hand, $S_{3b}$ is an efficient probe of the scalar interaction
since it requires only observations of the baryon distribution at
a fixed redshift.

We plot in Fig. 1 the functions $S_{3b,c}(\Omega _{m},\beta )$ obtained
through numerical integration. As anticipated, we find that $S_{3c}$
is close to the standard value 34/7 in the whole parameter range while
$S_{3b}$ deviates from it by more than 1\% for $\beta >0.1$, following
the approximate fit\begin{equation}
S_{3b}=\frac{34}{7}(1+0.6\beta ^{2})\Omega _{m}^{-0.0005(\beta ^{2})^{0.025}}.\label{eq:fits3b}\end{equation}
This result is almost independent of $\Omega _{m}$ (in fact the last
factor can be omitted) and $n$ and also independent of time (if \textbf{$\beta $}
is constant) and of scale: it shows therefore that $S_{3b}$ is a
\emph{direct test of the equivalence principle}. 

The analytical behavior (\ref{eq:s3bappr}) is relatively accurate
(error on $S_{3b}$$<1\%$) only for $\beta <0.2$. So far we assumed
$\beta _{b}=0$ but it is not difficult to see that, in the limit
$\Omega _{b}\ll \Omega _{c}$, Eqs. (\ref{eq:m}-\ref{eq:s3bappr}-\ref{eq:fits3b})
generalize to a finite $\beta _{b}$ by simply replacing $\beta ^{2}$
with $\beta _{c}(\beta _{c}-\beta _{b})$. Let us remark also that
although we performed the numerical integration with a dark energy
potential, the scalar field need not be the field responsible of the
accelerated expansion. The only condition on the potential for as
concern the validity of our numerical calculations is that the interaction
scale $\lambda $ be much larger than the astrophysical scale at which
the observations are carried out. For instance, the original Brans-Dicke
model, which does not give acceleration and where $V=0$, fulfills
this condition (we refer here to a Brans-Dicke model with a species-dependent
coupling as in Ref. \cite{dam}).

In \cite{review} the authors compiled an extensive list of present
constraints on the smoothed skewness $\hat{S}_{3}$ from angular and
redshift galaxy catalogs. Although several experiments quote values
of $\hat{S}_{3}$ with errors of 5-10\%, many results are clearly
not compatible with each other. Same scatter, if not larger, can be
seen in angular catalogs. This clearly points to the presence of systematic
errors, that are likely to reside in sampling and finite volume effects
and redshift distortions, so it is premature to perform a direct comparison
with data. However, analyses from larger redshift surveys like the
Sloan Digital Sky Survey (SDSS) promise to measure $\hat{S}_{3}$
at large scales with a precision of less than 10\% and perhaps down
to 1\% (see e.g. preliminary results in \cite{gazsdss,sdss} and,
for the 2dF survey, in \cite{croton}). At this level, SDSS might
detect the scalar interaction or put  a stringent upper limit to its
present value. 

It is however to be stressed that our calculations refer to the properties
of baryons, while observations deal with light, i.e. with the fraction
of baryons that collapsed in sufficiently bright galaxies. The relation
between the two populations is not well known, although at large scales,
where hydrodynamical effects and strong non-linearities are smeared
out, one does not expect significant segregation. To ascertain this
relation it will be necessary to perform $N$-body simulations with
broken equivalence, as in \cite{maccio}. On the other hand, it is
also possible to study objects that seem to trace with more accuracy
(or at least in a simpler way) the underlying baryon component, such
as Lyman-$\alpha $ clouds \cite{croft}. Errors less than 10\% in
the bispectrum at large scales are predicted in \cite{mandelbaum}
using a Lyman-$\alpha $ forest that simulate SDSS data.

Although models with non-Gaussian initial conditions, non-gravitational
effects or non-standard Friedman equations predict $S_{3}\not =34/7$
\cite{review,multa,sco}, they also predict a specific time and/or
scale dependence that make them distinguishable, at least in principle,
from a scalar interaction. Further information and can be gained by
the full bispectrum $B(\mathbf{k}_{1},\mathbf{k}_{2})=<\delta _{\mathbf{k}_{1}}\delta _{\mathbf{k}_{2}}\delta _{-\mathbf{k}_{1}-\mathbf{k}_{2}}>$,
rather than by the integrated skewness. In \cite{sco2} it has been
shown that the scale-dependence of the bispectrum may be of great
help in constraining primordial non-gaussianity. The behavior of the
bispectrum for the present model will be reported in subsequent work.
Forecoming large scale skewness data offer therefore the exciting
opportunity to test the equivalence principle in a realm inaccessible
to laboratory experiments.

\end{document}